%% file: isos.tex
\documentstyle[12pt,a41,epsfig]{article}
\def\Journal#1#2#3#4{{#1} {\bf #2}, #3 (#4)}

\def\NIMA{{\em Nucl. Instrum. Methods} A}
\def\NIMB{{\em Nucl. Instrum. Methods} B}

\def\NPBP{{\em Nucl. Phys.} B (Proc. Suppl.)}

\def\PLB{{\em Phys. Lett.}  B}
\def\PRL{\em Phys. Rev. Lett.}
\def\PRD{{\em Phys. Rev.} D}

\def\PRP{{\em Phys. Rep. }}
\def\EPC{{\em Europ. Phys. J.} C}

\def\PAN{\em Phys. Atom. Nucl.}

\def\RPP{\em Rep. Prog. Phys.}

\def\ASP{\em Astroparticle Physics}

\def\IJMP{\em Int. Journal of Modern Physics}
\def\JPG{\em Journal of Physics G}

\def\ra{\rightarrow}
\def\be{\begin{equation}}
\def\ee{\end{equation}}
\def\bea{\begin{eqnarray}}
\def\eea{\end{eqnarray}}

\newcommand{\gsim}{\mbox{$\stackrel{>}{\sim}$ }}

\newcommand{\expe}{experiment }

\newcommand{\exps}{experiments }

\newcommand{\obb}{0\mbox{$\nu\beta\beta$ decay}}

\newcommand{\nbb}{neutrinoless double beta decay }
\newcommand{\majo}{Majorana }

\newcommand{\mas}{Majorana neutrinos }

\newcommand{\delm}{\mbox{$\Delta m^2$} }

\newcommand{\me}{\mbox{$m_{\nu_e}$} }

\newcommand{\mmu}{\mbox{$m_{\nu_\mu}$} }

\newcommand{\nel}{\mbox{$\nu_e$} }
\newcommand{\nmu}{\mbox{$\nu_\mu$} }
\newcommand{\ntau}{\mbox{$\nu_\tau$} }

\newcommand{\munu}{\mbox{$\mu_{\nu}$} }
\newcommand{\mub}{\mbox{$\mu_B$} }
\newcommand{\ton}{\mbox{$T_{1/2}^{0\nu}$} }

\newcommand{\mamo}{magnetic moment}

\newcommand{\neu}{neutrino }
\newcommand{\neus}{neutrinos}

\newcommand{\ema}{\mbox{$\langle m_{\nu_e} \rangle$ }}

\newcommand{\gess}{\mbox{$^{76}Ge$ }}

\newcommand{\rehsa}{\mbox{$^{187}Re$ }}

\newcommand{\xehsd}{\mbox{$^{136}Xe$ }}

\newcommand{\cref}{\mbox{$^{51}Cr$ }}

\begin{document}
\sloppy
\begin{center}
\section*{Terrestrial neutrino mass searches}
\medskip
{\it K. Zuber$^a$}\\
$^a$ Lehrstuhl f\"ur Exp. Physik IV, Universit\"at Dortmund, 44221 Dortmund
\end{center}
\setcounter{section}{1}
\subsection{Introduction}
Neutrinos play a fundamental role in several fields of physics from cosmology down to
particle physics. Even more, the observation of a non-vanishing rest mass of neutrinos would
have a big impact on our present model of particle physics and might guide towards grand
unified
theories. Currently three evidences exist showing effects of massive neutrinos: the deficit in
solar neutrinos, the zenith angle dependence of atmospheric neutrinos and the excess events
observed by LSND. These effects are explained with the help of neutrino
oscillations, thus depending on \delm{} = $m_2^2 - m_1^2$, where $m_1,m_2$ are the \neu{} mass
eigenvalues and therefore are not absolute mass measurements. 
For a recent review on the physics of massive neutrinos see \cite{zub98}. 
\input massnel.tex
\input massmu.tex

\input masstau.tex
\input bb.tex

\input mamo.tex

\input ref.tex
\end{document}

%% file: massnel.tex
\subsection{Mass measurements of the electron neutrino}
The classical way to determine the mass of $\bar{\nel}$ (which is identical to $m_{\nu_e}$
assuming CPT invariance) is the
investigation of the
electron spectrum in beta decay.
A finite \neu mass will reduce the phase space and leads to a 
change of the shape
of the electron spectra.
In case several mass
eigenstates contribute, the total electron spectrum is given by a 
superposition
of the individual
contributions
\be
N(E) \propto F(E,Z) \cdot p \cdot E \cdot (Q-E) \cdot \sum^3_{i=1} 
\sqrt{(Q-E)^2 - m_i^2}
\mid U_{ei}^2
\mid 
\ee
where F(E,Z) is the Fermi-function, $m_i$ are the mass eigenvalues, $U_{ei}^2$ are
the mixing matrix elements connecting weak and mass eigenstates and $E,p$ are energy and momentum
of the emitted electron. The different involved $m_i$ produce kinks 
in the Kurie-plot 
where the size of the kinks is a measure
for the corresponding mixing
angle. This was discussed in connection with the now ruled out 17 keV
- \neu . A new sensitive search for kinks in the region 4-30 keV using
$^{63}$Ni
was done recently resulting in an overall upper limit of $ U_{e2}^2 < 10^{-3}$ \cite{hol99}.\\
Searches for an eV-\neu are done near the endpoint region of isotopes with low Q - values.
The preferred isotope under study is tritium, with an endpoint energy of about 18.6 keV.
By extracting o \neu mass limit out of their data, most \exps done in the past end up with
negative $m_{\nu}^2$ fit values,
which need
not to have a common origin. 
\begin{figure}
\begin{center}
\begin{tabular}{cc}
\epsfig{file=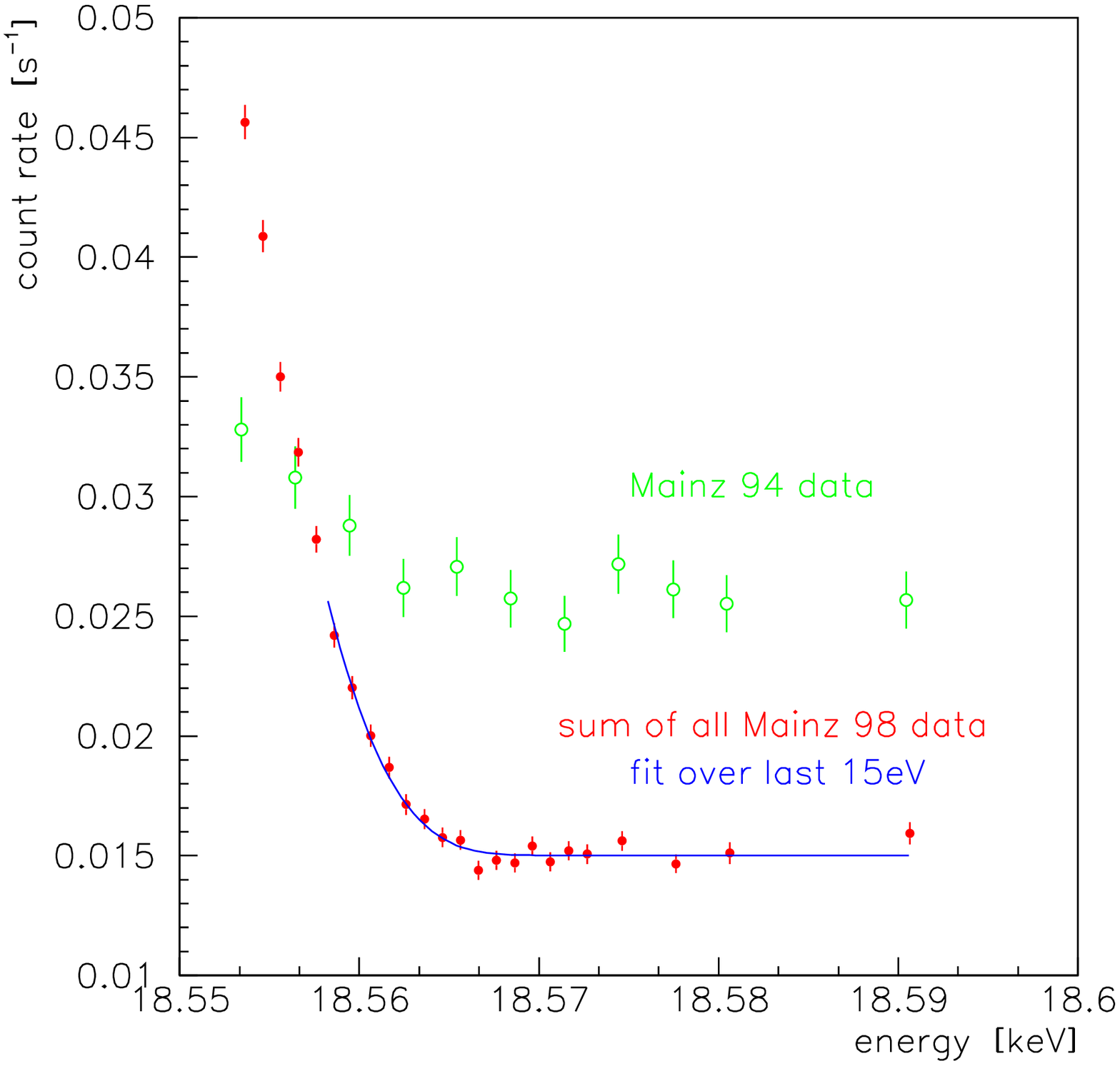,width=8cm,height=6cm} &
\epsfig{file=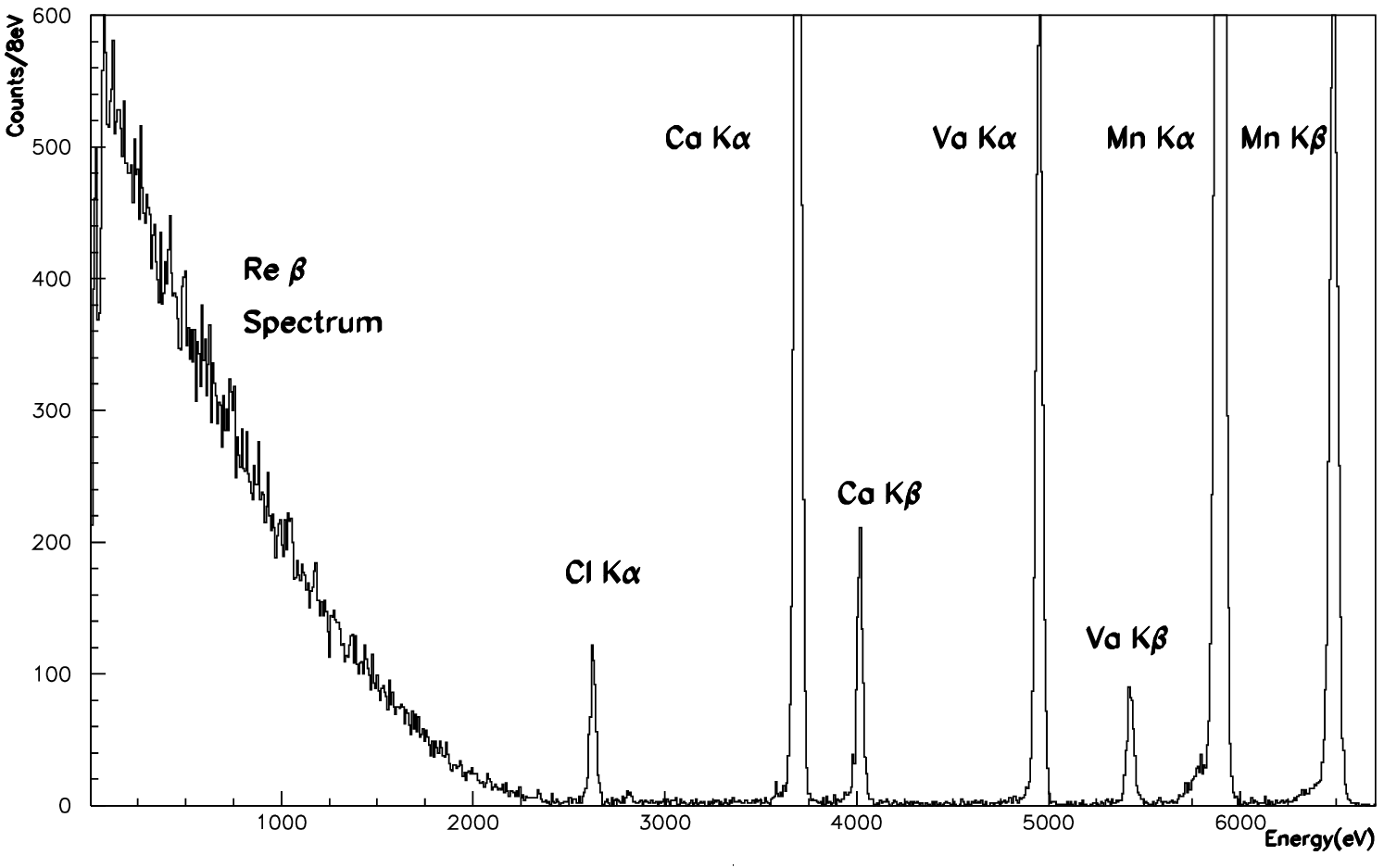,width=7cm,height=5cm}
\end{tabular}
\caption{left: Mainz 1998 electron spectrum near the endpoint of tritium decay. The
signal/background ratio is increased by a factor 
of 10 in comparison with the
1994 data. The Q-value of 18.574 keV is marking to the center of mass of the rotation-vibration 
excitations of the
molecular ground state of the daughter ion $^3HeT^+$. right: \rehsa $\beta$-spectrum obtained with a
cryogenic bolometer by the Genoa group. Calibration peaks can also be seen.}
\label{pic:mainz}
\end{center}
\end{figure}
For a detailed discussion of the \exps see \cite{hol92,ott95}.
While until
1990 mostly magnetic spectrometers were used
for the measurements, the new \exps in Mainz and Troitzk use electrostatic retarding
spectrometers \cite{lob85,pic92}. 
Fig.\ref{pic:mainz} shows the
present electron spectrum near the
endpoint as obtained with the Mainz spectrometer. 
The current obtained limits are 2.8 eV (95 \% CL) ($m_{\nu}^2 = - 3.7 \pm 5.3 (stat.) \pm 2.1 (sys.) eV^2$)
\cite{wei99} and 2.5 eV (95 \% CL)
($m_{\nu}^2 = - 1.9 \pm 3.4 (stat.) \pm 2.2 (sys.) eV^2$)
\cite{lob99} respectively. The final sensitivity should be around 2 eV.\\
Beside this, the Troitzk \expe observed excess counts in the
region of interest,
which can be described by a monoenergetic line a few eV below the endpoint. 
Even more, a semiannual modulation of the line position is observed \cite{lob99}. Clearly
further 
measurements are needed to investigate this effect. Considerations of building a new larger scale
version of such a spectrometer exist, to probe neutrino masses down below 1 eV.\\
A complementary strategy is followed by using cryogenic microcalorimeters. Because these
\exps measure the total energy released,
final state effects are not important. This method allows the investigation
of the $\beta$-decay of \rehsa, which has the lowest Q-value of all $\beta$-emitters (Q=2.67
keV). Furthermore the associated half-life measurement would be quite important, because
the \rehsa - $^{187}$Os pair is a well known cosmochronometer and a more precise half -
life measurement would sharpen the dating of events in the early universe like the formation of 
the solar system.
Cryogenic bolometers were build in form of metallic Re as well as AgReO$_4$ crystals and
$\beta$ - spectra (Fig.\ref{pic:mainz}) were measured \cite{gat99} \cite{ale99}, but at present 
the experiments
are not
giving any limits on \neu masses. Investigations to use this kind of technique also for
calorimetric measurements on tritium \cite{dep99} and on $^{163}$Ho \cite{meu98} are currently done.
Measuring accurately
branching ratios of atomic transitions or the internal bremsstrahlung spectrum in $^{163}$Ho
is interesting because this would result directly in a limit on \me{}.

%% file: massmu.tex
\subsection{Mass measurement of the muon \neu{}}
\begin{figure}
\begin{center}
\epsfig{file=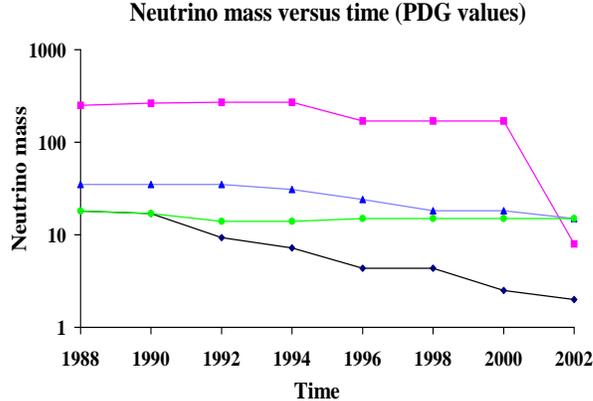,width=8cm,height=6cm}
\caption{Evolution of neutrino mass limits over the last 15 years using the Particle Data Group values.
Extrapolated values are given for 2000 and 2002. Electron neutrino limits are given for $\beta$-decay (black 
diamonds) and SN
1987A (green diamonds), for \nmu{} as triangles and \ntau as squares. As can be seen, the proposed measurement
of $m_{\nmu}$ at
the g-2 \expe{} would result in the largest factor obtained. The mass scale corresponds to eV (\nel), keV
(\nmu) and MeV (\ntau) respectively.}
\label{pic:pdg}
\end{center}
\end{figure}

The way to obtain limits on \mmu is given by the two-body decay of
the $\pi^+$.
A precise measurement of the muon momentum $p_{\mu}$ and
knowledge of $m_{\mu}$
and
$m_{\pi}$ is required. 
These measurement was done at the PSI resulting in a limit of \cite{ass96}
\be
\mmu^2 = (-0.016 \pm 0.023) MeV^2 \quad \ra \quad \mmu < 170 keV (90
\%CL)
\ee
A new idea looking for pion decay in flight using the g-2 storage ring at BNL has been proposed
recently \cite{cus99}. Because the g-2 ring would act as a high resolution spectrometer an
exploration
of \mmu down to 8 keV seems possible. Such a bound would have some far reaching consequences:
First of all it would be the largest step on any neutrino mass improvement within the last 20
years (Fig.\ref{pic:pdg}). Secondly it would bring any magnetic moment calculated within the 
standard model and associated with \nmu down to a
level of vanishing
astrophysical importance. Furthermore it would once and for all exclude that a possible 17 keV
mass eigenstate is the dominant contribution of \nmu . Possibly the largest impact is on
astrophysical topics. All bounds on \neu properties derived from stellar evolution are typically
valid for \neu masses below about 10 keV, so they would then apply for \nmu as well. For example, plasma
processes like $\gamma \ra \nu \bar{\nu}$ would contribute to stellar energy losses and significantly prohibit
helium ignition, unless the neutrino has a magnetic moment smaller than $\mu_{\nu} < 3 \cdot 10^{-12} \mu_B$
\cite{raf99} much more stringent than laboratory bounds. 

%% file: masstau.tex
\subsection{Mass measurement of the tau \neu{}}
The present knowledge of the mass of \ntau stems from measurements with
ARGUS, CLEO, OPAL, DELPHI and ALEPH (see \cite{pas97}).
Practically all \exps use the $\tau$-decay into five charged pions
$\tau \ra \ntau + 5\pi^{\pm} (\pi^0)$
with a branching ratio of BR = ($9.7 \pm 0.7) \cdot 10^{-4}$. To increase the
statistics CLEO, OPAL, DELPHI
and ALEPH
extended their search by including the 3 $\pi$ decay mode. But even with the 
disfavoured statistics,
the 5 prong-decay is much more sensitive, because the mass of the 
hadronic system peaks at about 1.6 
GeV, while the 3-prong system is dominated by the $a_1$ resonance at 
1.23 GeV. While ARGUS obtained their limit by investigating the invariant mass of the 
5 $\pi$-system, ALEPH, CLEO and OPAL 
performed a two-dimensional analysis by including the energy of 
the hadronic system.
The most 
stringent one is given by ALEPH \cite{bar98}

%% file: bb.tex
\subsection{Double beta decay}
\label{ch:cha41}
\begin{figure}[bht]
\begin{center}
\epsfig{file=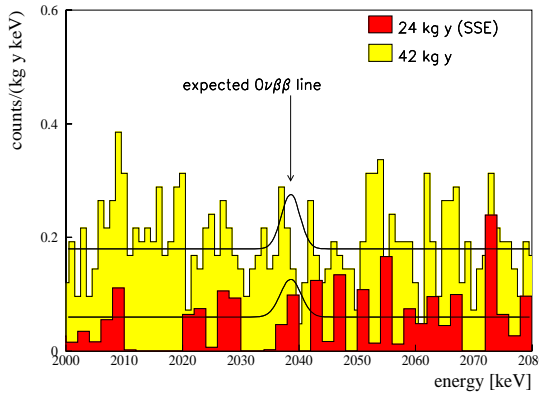,width=7cm,height=5cm}
\caption{Observed sum energy spectrum of the electrons around the expected \obb{} line position
obtained by the
Heidelberg-Moscow experiment. No signal peak is seen. The two
different spectra correspond to data sets with (black) and 
without (grey) pulse
shape discrimination.}
\label{pic:heimo}
\end{center}
\end{figure}

The most promising way to distinguish between Dirac and \mas is \nbb (\obb{})
\be
(Z,A) \ra (Z+2,A) + 2 e^-  \quad (\Delta L =2)
\ee
only possible if 
\neus are massive \majo particles.
The measured quantity is called effective \majo \neu mass \ema and given by
\be
\label{eq:ema}\ema = \mid \sum_i U_{ei}^2 \eta_i m_i \mid
\ee
with the relative CP-phases $\eta_i = \pm 1$, $U_{ei}$ as the mixing
matrix elements and
$m_i$ as the
corresponding mass eigenvalues. 
From the experimental point, the evidence for \obb is a peak in the sum energy
spectrum of the electrons 
at the
Q-value of the involved transition. 
The best limit is coming from the Heidelberg-Moscow \expe resulting in a 
bound of \cite{bau99}
(Fig.\ref{pic:heimo}) 
\be
\label{eq:thalb}
\ton > 5.7 \cdot 10^{25} y \ra \ema < 0.2 eV \quad (90 \% CL)
\ee 
having a sensitivity of $\ton > 1.6 \cdot 10^{25} y$. 
Eq.(\ref{eq:ema}) has to be modified in case of heavy \neus ($m_{\nu}
\gsim $1 MeV). For such heavy \neus the mass can no longer be neglected in the
\neu propagator resulting in an A-dependent
contribution
\be
\ema =  \mid \sum_{i=1,light}^N U^2_{ei} m_i + \sum_{h=1,heavy}^M F (m_h,A) 
U^2_{eh} m_h  \mid
\ee
By comparing these limits for isotopes with different atomic mass, 
interesting limits
on the mixing angles and \ntau parameters for an MeV \ntau 
can be obtained \cite{hal83,zub97}.
\paragraph{Future}
Several upgrades are planned to improve the existing half-life limits, only three are mentioned here, for
details see \cite{zub98}.
The next to come is NEMO-3, a giant TPC using double beta emitters up to 10 kg in form of thin foils,
which should start operation in 2000.
Even more ambitious would be the usage of
large amounts of materials (in the order of several hundred kg to tons)
like enriched \xehsd added to
scintillators
\cite{rag94}, 750 kg $TeO_2$ in form of cryogenic bolometers (CUORE) \cite{fio98} or a
huge cryostat containing several hundred detectors of
enriched \gess with a total mass of 1 ton (GENIUS) \cite{kla98}.

%% file: mamo.tex
\subsection{Magnetic moment of the neutrino}
Another possibility to check the \neu character and mass is the search for its \mamo{}.
In the case of Dirac \neus{}, it can be shown that \neus can have a \mamo
due to loop diagrams which is proportional to
their mass and is given by \cite{lee77,mar77}
\be
\munu = \frac{3 G_F e}{8 \sqrt{2} \pi^2} m_{\nu} = 3.2 \cdot 10^{-19} (\frac{m_\nu}{eV}) \mub
\ee
In case of \neu masses in the eV-range, this is far to small to be observed
and to have any significant
effects in
astrophysics. Nevertheless
there exist GUT-models, which are able to increase the \mamo without increasing the mass
\cite{pal92}. However
\majo \neus still have a vanishing static moment because of CPT-invariance.
The existence of diagonal terms in the \mamo matrix would therefore prove 
the
Dirac-character of \neus.
Non-diagonal terms in the moment matrix are possible for both types of \neus
allowing transition moments of the form \nel - $\bar{\nu}_\mu$.\\ 
Limits on magnetic moments arise from \nel $e$ - scattering \exps and 
astrophysical considerations. The 
differential cross section for \nel $e$ -
scattering in presence of a \mamo is given by
\bea
\frac{d \sigma}{dT} = \frac{G_F^2 m_e}{2 \pi}
[(g_V + x+g_A)^2 +
(g_V + x- g_A)^2 (1-\frac{T}{E_\nu})^2 \\
+ (g_A^2 -
(x+g_V)^2)\frac{m_e T}{E_\nu^2}] + \frac{\pi \alpha^2 \munu^2}{m_e^2}
\frac{1-T/E_\nu}{T}
\eea
where T is the kinetic energy of the recoiling electron and 
$x$ denotes the \neu form factor related to its square charge radius $\langle r^2 \rangle$
\be 
x=\frac{2 m_W^2 }{3} \langle r^2 \rangle sin^2 \theta_W \quad x \ra -x \quad for 
\quad \bar{\nel}
\ee
The contribution associated with the charge radius can be neglected in the case $\mu_\nu
\stackrel{>}{\sim} 10^{-11} \mub$. 
As can be seen, the largest effect of a \mamo can be observed in the low
energy region, and because of
destructive interference
of the electroweak terms, searches with antineutrinos would be preferred. The obvious sources
are therefore nuclear
reactors. Experiments done so far give limits of \munu{} $<1.8 \cdot 10^{-10} \mu_B$ (\nel), \munu $<7.4
\cdot
10^{-10} \mu_B$ (\nmu) and \munu{} $<5.4 \cdot 10^{-7} \mu_B$ (\ntau). Also bounds for a \mamo of a sterile
neutrino,
discussed in
more detail later, can be obtained from a Primakoff like conversion in $\nu$N scattering 
if there is a mixing with \nmu. \\
Astrophysical limits are somewhat more stringent but also more model dependent. 
To improve the experimental situation new experiments are taking data or are under construction. The most
advanced is the
MUNU \expe \cite{ams97} currently running at
the Bugey
reactor. It consists of a 1 m$^3$ TPC loaded with CF$_4$ under a pressure of 5 bar. The usage 
of a TPC will not only
allow to measure the electron energy but for the first time in such \exps also the 
scattering angle, making the
reconstruction of the neutrino energy possible. 
In case of no \mamo the expected count rate is 9.5 per 
day increasing to 13.4 per day if 
$\munu= 10^{-10} \mub$ for an energy threshold of 500 keV. The estimated 
background is 6 events per day. The expected
sensitivity level is down to $\munu = 3 \cdot 10^{-11} \mub$ . The usage 
of a low background Ge-NaI
spectrometer in a shallow depth near a reactor has
also been considered \cite{bed97}. The usage of large low-level detectors
with a low-energy threshold
of a few keV in underground laboratories is also under investigation. The reactor 
would be replaced
by a strong $\beta$-source. Calculations for a scenario of a 1-5 MCi $^{147}$Pm 
source (endpoint
energy of 234.7 keV) in combination with a 100 kg low-level NaI(Tl) detector with a 
threshold of about 2
keV
can be found in \cite{bar96}. Also using a $^{51}$Cr source within the BOREXINO experiment will allow
to put stringent limits on $\munu$.

%% file: isos.bbl
\begin{thebibliography}{99}
\bibitem{zub98} K. Zuber, \Journal{\PRP}{305}{295}{1998}
\bibitem{hol99} E. Holzschuh et al.,\Journal{\PLB}{451}{247}{1999}
\bibitem{hol92} E. Holzschuh, \Journal{\RPP}{55}{1035}{1992}
\bibitem{ott95} E.W. Otten, \Journal{\NPBP}{38}{26}{1995}
\bibitem{lob85} V.M. Lobashev et al., \Journal{\NIMA}{240}{305}{1985}
\bibitem{pic92} A. Picard et al., \Journal{\NIMB}{63}{345}{1992}
\bibitem{wei99} C. Weinheimer et al., \Journal{\PLB}{460}{219}{1999}
\bibitem{lob99} V.M. Lobashev et al., \Journal{\PLB}{460}{227}{1999}
\bibitem{gat99} F. Gatti et al., \Journal{Nature}{397}{137}{1999}
\bibitem{ale99} A. Alessandrello et al., \Journal{\PLB}{457}{253}{1999}
\bibitem{dep99} D. Deptuck, private communication
\bibitem{meu98} P. Meunier, \Journal{\NPBP}{66}{207}{1998}
\bibitem{ass96} K. Assamagan et al., \Journal{\PRD}{53}{6065}{1996}
\bibitem{cus99} P. Cushman , K. Jungmann, private communication
\bibitem{raf99} G. Raffelt, Prep. hep-ph/9903472, to appear in ARNPS Vol.49
\bibitem{pas97} L. Passalacqua, \Journal{\NPBP}{55C}{435}{1997}
\bibitem{bar98} R. Barate et al., \Journal{\EPC}{2}{395}{1998}
\bibitem{bau99} L. Baudis et al., \Journal{\PRL}{83}{41}{1999}
\bibitem{hal83} A.Halprin, S.T.Petcov, S.P.Rosen, \Journal{\PLB}{125}{335}{1983}
\bibitem{zub97} K. Zuber, \Journal{\PRD}{56}{1816}{1997}
\bibitem{rag94} R.S. Raghavan, \Journal{\PRL}{72}{1411}{1994}
\bibitem{fio98} E. Fiorini, private communication
\bibitem{kla98} H.V. Klapdor-Kleingrothaus, J. Hellmig, M. Hirsch, 
\Journal{\JPG}{24}{483}{1998}
\bibitem{lee77} B.W. Lee, R.E. Shrock, \Journal{\PRD}{16}{1444}{1977}
\bibitem{mar77} W.J. Marciano, A.I. Sanda, \Journal{\PLB}{67}{303}{1977}
\bibitem{pal92}P.B. Pal, \Journal{\IJMP}{A7}{5387}{1992}
\bibitem{kra90} D. Krakauer et al., \Journal{\PLB}{252}{171}{1990}
\bibitem{abe87} K. Abe et al., \Journal{\PRL}{58}{636}{1987}
\bibitem{kim88} C.S. Kim, W.J. Marciano, \Journal{\PRD}{37}{1368}{1988}
\bibitem{ams97}C. Amsler et al., \Journal{\NIMA}{396}{115}{1997}
\bibitem{bed97} A.G. Beda et al., \Journal{\PAN}{61}{66}{1998}
\bibitem{bar96} I. Barabanov et al., \Journal{\ASP}{5}{159}{1996}
\end{thebibliography}
